# Design for Sensing and Digitalisation (DSD): A Modern Approach to Engineering Design

Daniel N. Wilke[a,b]

[a]Department of Mechanical Engineering and Aeronautical Engineering, University of Pretoria, Pretoria, South Africa

[b]School of Mechanical, Industrial and Aeronautical Engineering, University of the Witwatersrand, Johannesburg, South Africa

*corresponding author email: daniel.wilke@wits.ac.za

**Abstract**

This paper introduces Design for Sensing and Digitalisation (DSD), a new engineering design paradigm that integrates sensor technology for digitisation and digitalisation from the earliest stages of the design process. Unlike traditional methodologies that treat sensing as an afterthought, DSD emphasises sensor integration, signal path optimisation, and real-time data utilisation as core design principles. The paper outlines DSD's key principles, discusses its role in enabling digital twin technology, and argues for its importance in modern engineering education. By adopting DSD, engineers can create more intelligent and adaptable systems that leverage real-time data for continuous design iteration, operational optimisation and data-driven predictive maintenance.

*Keywords*: Digitisation, Digitalisation, Sensing, Design, Design for Sensing and Digitalisation (DSD), Industry 4.0, Digital Twins, Sensor Integration, Signal Path Optimisation, Optimal Sensor Placement, Engineering Design, Predictive Maintenance, Real-time Monitoring, Engineering Education, Smart Manufacturing

## 1. INTRODUCTION

The digital transformation of industry has fundamentally changed how engineered systems are designed, operated, and maintained. While traditional design methodologies like Design for Manufacturing (DFM) [1] and Design for Maintenance (DFMaint) [2] remain valuable, they fail to address a critical requirement of modern engineering, which is the planning, design and integration of sensing capabilities and data acquisition systems. This paper introduces Design for Sensing and Digitalisation (DSD), a design methodology that places sensor integration and data acquisition at the core of the engineering design process.

DSD addresses three key challenges in modern engineering design. First, the growing demand for real-time monitoring and optimisation requires seamless sensor integration, yet current methodologies typically treat sensing as an afterthought. Second, digital twin technology necessitates high-fidelity data collection, which can only be achieved through carefully planned sensor placement [3] and signal path optimisation. Third, the increasing complexity of engineered systems requires sophisticated predictive maintenance capabilities that depend on comprehensive sensor coverage and reliable data acquisition.

By incorporating sensing and digitalisation considerations from the earliest stages of design, DSD enables engineers to create systems inherently capable of generating the high-quality data needed for modern applications. This approach enhances system performance through real-time monitoring and facilitates predictive maintenance, continuous design iteration and operational optimisation, ultimately leading to more reliable and efficient engineered systems.

### 1.1 The Imperative of Digitalisation

The Internet of Things (IoT) [4] and digital twins [5] demand a fundamental shift in how we approach the engineering design of physical assets and processes. While digital twins and digital processes offer unprecedented insights through real-time monitoring and analysis, their effectiveness hinges entirely on the quality and completeness of the sensor data they receive. This creates a critical challenge: sensor integration must be considered from the earliest design stages, not as an afterthought. Furthermore, modern systems require data collection and optimised sensor networks that ensure comprehensive observability of a system's state or inference, using virtual sensing while minimising noise and interference.

The challenge extends beyond new designs. Legacy systems require sensor retrofitting, often in environments where existing structures constrain optimal sensor placement. These constraints frequently result in compromised data quality and incomplete system observability, limiting the potential of digital twin implementation. This underscores the need for a design methodology that considers sensing capabilities as fundamental to the design rather than an add-on feature.

### 1.2 Limitations of Traditional Design Methodologies

While effective in their respective domains, traditional design methodologies often fall short when confronted with the complexities of digitisation. DFM [1], for instance, focuses on optimising a design for ease of manufacturing, while DFMaint [2] emphasises reducing maintenance costs and downtime. However, neither methodology explicitly addresses the crucial data acquisition and utilisation aspect central to the digital age.

Neither addresses three fundamental challenges:

1. Sensor placement and signal path optimisation are typically afterthoughts, leading to suboptimal data collection and potential blind spots in system monitoring.

2. System observability is rarely considered during initial design phases, resulting in situations where critical parameters cannot be effectively measured or monitored.

3. The lack of integrated sensor simulation and virtual calibration capabilities [7,8] means that sensor network effectiveness cannot be verified before physical implementation.

These limitations create a significant gap between traditional design approaches and the requirements of modern digital twin implementations. When sensing and digitalisation are not core design considerations, the resulting systems often require costly modifications or deliver incomplete data, compromising the effectiveness of digital twin applications and limiting opportunities for data-driven operational optimisation, design iteration and predictive maintenance.

## 2. DESIGN FOR SENSING AND DIGITALISATION (DSD): ADDRESSING MODERN DESIGN CHALLENGES

DSD extends virtual calibration [7,8] beyond its traditional role of quantifying parameter estimation uncertainty from experimental data. Instead, DSD empowers designers to proactively control this uncertainty by incorporating sensor networks and system observability as fundamental to system design. Through simulation-driven sensor placement [3] and sensor network optimisation [3], DSD ensures that critical system parameters, states, and conditions are observable by design rather than leaving observability to chance or post-design modifications.

This proactive approach naturally incorporates traditional design considerations like manufacturing [1], maintenance [2], and reliability, as the designer can now ensure that key parameters affecting these aspects are continuously monitored and quantified. By putting designers in control of system observability from the outset, DSD transforms the relationship between physical systems and their digital twins, ensuring that the data needed for effective system monitoring, optimisation, and maintenance is available by design rather than by circumstance.

DSD fundamentally reimagines the design process by integrating sensing and digitisation requirements from conception. This paradigm addresses four critical aspects:

### 2.1 System State Observability by Design

DSD ensures system state observability through:

1. Early identification of critical parameters requiring monitoring.
2. Mathematical modelling to confirm complete system state observability before physical implementation [8].
3. Optimise sensor placement to maximise information capture while minimising sensor count [3].
4. Integration of virtual sensing capabilities where physical sensors are impractical or cost-prohibitive.

DSD makes system state observability a fundamental design requirement by identifying critical parameters and validating observability through mathematical modelling before implementation. This approach ensures comprehensive system monitoring while minimising sensor counts through optimisation-driven sensor placement and virtual sensing strategies. The integration of observability analysis in the design phase enables efficient monitoring of system states throughout the product lifecycle, supporting immediate operational needs and long-term system optimisation.

### 2.2 Signal Path Optimisation

The paradigm prioritises signal integrity through:

1. Design-stage simulation of sensor networks to identify potential interference patterns [6].
2. Optimisation of physical layouts to minimise signal degradation.
3. Strategic placement of signal conditioning and processing elements [6].
4. Implementation of noise reduction strategies as core design features.

Signal path optimisation in DSD treats signal quality as a primary design constraint through simulation-driven design, optimisation of sensor networks, and strategic component placement. Addressing signal integrity and interference issues during design ensures high-fidelity data collection while reducing implementation costs and system complexity. The resulting framework provides reliable data for effective system monitoring throughout the product lifecycle.

### 2.3 Digital Twins For Smart Design & Redesign

DSD integrates digital twins for design and redesign by:

1. Naturally extending redesign through collected data to enable smart redesign.
2. Creating high-fidelity sensor networks that ensure comprehensive data collection.
3. Establishing verified and validated simulation models during the design phase.
4. Enabling real-time operational comparisons between predicted and actual system behaviour.
5. Facilitating continuous model refinement through operational data [9].

DSD creates a natural feedback loop where operational data directly informs design modifications and redesign iterations, providing a foundation for data-driven smart design & redesign, predictive maintenance, and system optimisation throughout the product and process lifecycle [5].

### 2.4 Adaptive Systems

DSD naturally enable the design of adaptive systems

1. Real-time system identification enables continuous model updating and refinement based on operational data, allowing systems to adapt to changing conditions and ageing effects.
2. Optimised sensor networks enable automated detection of system changes and anomalies, facilitating rapid response to evolving operational requirements.
3. Digital twin integration allows for virtual testing of adaptation strategies before physical implementation, reducing risks associated with system modifications.
4. Multi-fidelity modelling frameworks capture rapid dynamic changes and long-term system evolution, enabling adaptive responses across different time scales.

DSD enables the development of adaptive systems that adjust to changing operating conditions and requirements in real-time. This is particularly valuable in dynamic environments where flexibility is essential.

## 3. DSD: A CORE FRAMEWORK FOR MODERN ENGINEERING EDUCATION

DSD serves as both a design methodology and an educational paradigm, bridging the gap between traditional engineering disciplines and the demands of Industry 4.0.

DSD provides a natural framework for integrating traditionally separate engineering disciplines and modern digital competencies into a cohesive curriculum. By centring design around sensing and digitalisation, DSD inherently connects fundamental engineering principles with essential digital age skills:

### 3.1 Integration of Core Competencies

DSD naturally incorporates key modern engineering skills through:

1. Statistical analysis for sensor network design and data interpretation.
2. Programming and data science for digital twin implementation.
3. Machine learning for system optimisation and predictive analytics.
4. Computer-aided engineering (CAE) for system modelling and simulation.
5. Optimisation techniques for sensor placement and signal path design.

DSD provides a natural framework for developing modern engineering competencies by integrating statistical analysis, programming, and data science into the design process. Training in data analytics and AI techniques enables students to leverage the vast amounts of data generated by sensor-integrated systems. Through sensor network design and digital twin implementation, students engage with computer-aided engineering, machine learning, and optimisation techniques in practical applications. This integrated approach ensures that essential digital skills are developed within the context of engineering design rather than as isolated technical capabilities.

### 3.2 Bridging Traditional and Modern Engineering

The DSD framework seamlessly connects traditional design considerations [10] with digital capabilities:

1. Manufacturing constraints inform sensor placement and integration.
2. Maintenance requirements drive monitoring strategy development.
3. Reliability engineering guides sensor network redundancy.
4. System dynamics influence sensing frequency and placement.
5. Material selection considers both structural and sensing requirements.

The DSD framework integrates traditional engineering design with digital capabilities by considering sensing and digitalisation in every design decision. Manufacturing constraints and maintenance requirements shape sensor integration and monitoring strategies, while system dynamics and reliability considerations guide sensor network design. This approach ensures that all design decisions, from material selection to structural layout, address the physical system, flow of information and digitalisation of information requirements of modern engineered systems.

### 3.3 Integrated Approach

A DSD-based curriculum emphasises an integrated approach through:

1. Project-based learning combining physical design and digital implementation.
2. Integration of industry-standard tools and practices.
3. Real-world case studies demonstrating DSD principles.
4. Team-based projects reflecting modern engineering workflows.
5. Continuous feedback between design and implementation phases.

This integrated approach prepares engineers to address complex system design challenges while naturally accommodating traditional design considerations within a modern digital framework.

## 4. CONCLUSION

The Design for Sensing and Digitalisation (DSD) paradigm shifts sensor integration and signal optimisation from secondary considerations to core design principles. DSD ensures comprehensive monitoring and optimisation from the outset by prioritising system observability through optimised sensor networks, enhancing signal integrity with strategic design choices, and integrating digital twins for adaptive responses. This proactive approach improves data quality, enhances system performance, and embeds maintenance, manufacturing, and reliability considerations throughout the system lifecycle.

DSD also supports modern engineering education by integrating digital competencies with traditional principles. Hands-on experience with sensor networks, digital twins, and data-driven optimisation equips students with the skills necessary for Industry 4.0, reinforcing a cycle of improved design practices and capabilities.

The result is a robust and adaptable methodology that meets the demands of digital engineering while preparing engineers to drive industry transformation. DSD is a natural, practical design approach and educational framework that fosters continuous system improvement and technological advancement.

## 5. ACKNOWLEDGEMENTS

The authors acknowledge using artificial intelligence tools to assist in refining and expanding the language of initial manuscript drafts. All conceptual, analytical, and technical content remains the authors' original work, and AI tools were employed solely to support clarity and consistency in writing.

**REFERENCES**

[1] Anderson, D.M. (2020). "Design for Manufacturability: How to Use Concurrent Engineering to Develop Low-Cost, High-Quality Products for Lean Production Rapidly." CRC Press.

[2] Gullo, L. J., & Dixon, J. (Eds.). (2021). Design for Maintainability. John Wiley & Sons, Ltd.

[3] Chae, Y., and D. N. Wilke. 2017. "Heuristic Linear Algebraic Rank-Variance Formulation and Solution Approach for Efficient Sensor Placement." *Engineering Structures* 153: 717–731.

[4] Madakam, S., Ramaswamy, R. and Tripathi, S. (2015) Internet of Things (IoT): A Literature Review. Journal of Computer and Communications, 3, 164-173.

[5] Wilke, D. N. (2022). "Digital Twins for Physical Asset Lifecycle Management." In *Digital Twins: Basics and Applications*, 13–26.

[6] Long, D., Hong, X., & Dong, S. (2006). Signal-path driven partition and placement for analog circuit. *Asia and South Pacific Conference on Design Automation, 2006*, Yokohama, Japan, pp. 6.

[7] Wilke, D. N., Kok, S., & Groenwold, A. A. (2010). The application of gradient-only optimization methods for problems discretized using non-constant methods. Structural and Multidisciplinary Optimization, 40, 433-451.

[8] Ben Turkia, S., Wilke, D. N., Pizette, P., Govender, N., & Abriak, N. E. (2019). Benefits of virtual calibration for discrete element parameter estimation from bulk experiments. Granular Matter, 21, 1–16.

[9] Booyse, W., Wilke, D. N., & Heyns, S. (2020). Deep digital twins for detection, diagnostics, and prognostics. Mechanical Systems and Signal Processing, 140, 106612.

[10] Wilke, D. N. 2023. "A Personal Perspective on the Engineering Design Process and Design Qualification." Medium, Month Day. https://medium.com/@wilkedn/a-personal-perspective-on-the-engineering-design-process-and-design-qualification-d6352a5cca55.